\documentclass[aps,prb,twocolumn,floats,showpacs]{revtex4}
\usepackage{bm,graphicx,epsf}

\newcommand{\be}{\begin{equation}}
\newcommand{\ee}{\end{equation}}
\newcommand{\bea}{\begin{eqnarray}}
\newcommand{\eea}{\end{eqnarray}}
\newcommand{\phrl}[1]{Phys.~Rev.~Lett. {\bf #1}}
\newcommand{\phrb}[1]{Phys.~Rev.~B {\bf #1}}

\newcommand{\bib}{\bibitem}
\newcommand{\lb}{\left[}
\newcommand{\rb}{\right]}

\newcommand{\G}{{\cal G}}
\renewcommand{\r}{{\bf r}}
\renewcommand{\j}{{\bf j}}

\newcommand{\q}{{\bf q}}
\renewcommand{\k}{{\bf k}}

\begin{document}

\title{Frequency- and transverse wave-vector-dependent spin Hall conductivity in two-dimensional electron gas with disorder}
\author{Sudhansu S. Mandal and Ankur Sensharma\cite{Ankur}
}
\affiliation{Theoretical Physics Department, Indian Association for the 
   Cultivation of Science, Jadavpur, Kolkata 700 032, India}

\date{\today}


\begin{abstract}
We  determine wave number $q$ and frequency $\omega$ 
dependent spin Hall conductivity $\sigma_{yx}^s(q , \omega)$ for a disordered  
two dimensional electron system with Rashba spin orbit interaction
when $\q$ is {\it transverse} to the electric field.
Both the conventional definition of spin current and its new definition which 
takes care of the conservation of spins, have been considered. 
The spin Hall conductivitivities for both of these definitions are qualitatively similar.
$\sigma_{yx}^s(q , \omega)$ is zero at $q=0,\,\omega =0$ and is maximum at $q=0$ and at
small but finite $\omega$ whose value depends on different parameters of the system. 
Interestingly for $\omega \to 0$, $\sigma_{yx}^s(q)$ resonates
when $\Lambda \simeq L_{so}$ which are the wavelength $(\Lambda = 2\pi/q)$
of the electric field's spatial variation  and the length for one cycle
of spin precession respectively. 
The sign of the out-of-plane component of the electrons' spin flips
when the sign of electric field changes due to its spatial variation
along transverse direction. It changes the mode of spin 
precession from clockwise to anti-clockwise or {\it vice versa}
and consequently a finite spin Hall current flows in the bulk of the system.

\end{abstract}

\maketitle

\section{Introduction}

One of the primary goals today in spin based electronics \cite{spintronics}
is the generation of spin current. 
Recent realization of spin-Hall effect (SHE) \cite{expt1,expt2,expt3}
in semiconductor
systems is certainly a very significant achievement in this direction. 
This is a phenomenon for electrical generation of spin: a charge
current along its transverse direction induces a spin current whose
polarization is perpendicular to the plane formed by these two currents. 
This phenomenon was predicted \cite{shallth1,shallth2,shallth3}
long back  due to the spin-asymmetric `skew-scattering' mechanism
and spin dependent `side-jump' mechanism 
which are collectively called {\em extrinsic} mechanism because the
spin-orbit interaction (SOI) is disordered in this case. This mechanism is also
responsible for anomalous Hall effect (AHE) \cite{ahe2,ahe3}
in ferromagnets. However uniform
(pure) SOI which is {\em intrinsic} \cite{ahe1}, also causes AHE. 
Similar {\em intrinsic} mechanism due to the SOI 
in hole-doped semiconductors \cite{int1} and two 
dimensional electron gas \cite{int2}
in semiconductor heterostructures have been predicted
to give rise to nondissipative spin Hall conductivity (SHC).

The spin accumulation observed in $n$-doped GaAs \cite{expt1}
is believed to be extrinsic
in origin because of small spin accumulation and its
directional independence on the electric field. 
On the other hand spin accumulation
in two dimensional hole gas (2DHG) is large \cite{expt2}
and hence is suggested to be intrinsic in origin. 
These experiments do not measure the spin voltage or the spin current, however
the technique developed in observing charge accumulation \cite{ishall}
at the transverse edge
due to spin current which is called inverse SHE, could be useful to measure
spin Hall current.
Nevertheless estimated SHC from the observation of spin
accumulation \cite{expt1} in $n$-doped GaAs is in good agreement with 
the calculated \cite{ext1,ext2} SHC in extrinsic mechanisms. 
An effective two band cubic (in momentum) Rashba model 
describes the electronic states in 2DHG well. 
Even in presence of disorder, the SHC in 2DHG has nonzero \cite{2dhg}
intrinsic contribution.

The responsible mechanism for SHE in two dimensional electron gas
(2DEG) which we study here is particularly not clear yet.
The electronic states in the 2DEG formed in semiconductor heterostructures
can be well described by the Hamiltonian  
\be
 H_0 = \frac{\k^2}{2m}\sigma_0 + \lambda (k_y\sigma_1 - k_x \sigma_2) \label{Hamiltonian}\, ,
\ee
where $\k$ and $m$ are the momentum and mass of the electrons respectively,
$\lambda$ is the Rashba spin orbit coupling strength \cite{Rashba},
$\sigma_0$ is the unit matrix, and $\sigma_i$ are the Pauli matrices. (We have set the unit $\hbar =c =1$.)
Sinova {\it et al} \cite{int2} predicted universal SHC $\sigma_{yx}^s = e/8\pi$ for
such systems using conventional definition of spin current 
$\hat{\j}^\alpha = \frac{1}{2}\left\{ s_\alpha,\, \hat{\bm{v}}_\k\right\}$ where
group velocity $\hat{\bm{v}}_\k = \bm{\nabla}_\k H_0$ and 
$s_\alpha = \sigma_\alpha /2 $ is the
 $\alpha$-th $(\alpha=1-3)$ component of spin. However after a prolonged
debate \cite{dis1,dis2,dis2a,dis3,dis4,dis5,num1,num2,Rashba2,sshall,Nagaosa}, 
the consensus arising from various methods of calculations is that 
$\sigma_{yx}^s = 0$ in presence of disorder, no matter what its strength is.
These studies include calculation of vertex correction \cite{dis1}
in Born approximation,
using Keldysh formalism \cite{dis2,dis2a,Nagaosa}
for any value of lifetime $\tau$, using Kubo
formula analytically \cite{dis3,dis4,sshall} and numerically \cite{num1,num2}, 
and Boltzmann transport equation approach \cite{dis5}. 
The equation of motion for spin projected on the plane is
$\partial_t (\sigma_1 ,\sigma_2) = -4m\lambda (\hat{j}^3_x, \hat{j}^3_y)$.
A very unique feature \cite{dis3,Rashba2} of the linear Rashba model is that 
$\partial_t \sigma_2$ is proportional to $\hat{j}^3_y$.
 It suggests zero spin Hall current in the steady state \cite{dis3}. 
This simple argument
describes the vanishing $\sigma_{yx}^s$ for such systems. A similar argument
also describes spin-spin-Hall current \cite{sshall}
since $j^1_x= -j^2_y$ for steady state
derivable from the equation
$\partial_t \sigma_3 = 4m\lambda(\hat{j}^1_x +\hat{j}^2_y)$.
All these results are obtained from the above conventional definition of 
$\hat{\j}^\alpha$ which is not conserved. The new definition of 
conserved spin current proposed by Shi {\it et al.} \cite{cspin}
gives rise to vanishing total $\sigma_{yx}^s$
for short-ranged $\delta (\r)$ impurity potential and for long-ranged
potential upto first order Born approximation \cite{Nagaosa}.

Is then the `intrinsic' mechanism really absent for spin Hall effect in 2DEG?
In a disordered 2DEG, an in-plane applied magnetic field may lead to the nonvanishing
intrinsic SHC \cite{Zeeman} due to Zeeman coupling. Further the interplay
of Zeeman coupling with different spin-orbit interactions may also lead to finite SHC \cite{Zeeman2}
in a pure system.
In this paper, we calculate SHC 
 using Kubo formula at finite frequency $\omega$
and momentum $\q$ {\it transverse} to the applied electric field within the 
intrinsic mechanism in a disordered 2DEG with no applied magnetic field. 
We find that even in the static limit, SHC is nonvanishing and hence  
the presence of `intrinsic' mechanism for spin Hall effect in 2DEG is demonstrated.

The paper is organized as follows. In the next section, we calculate frequency and transverse
momentum dependent SHC in a disordered 2DEG with Rashba SOI using Kubo formula with the 
conventional definition of spin-current.
The contribution of spin torque to the SHC
is also calculated and this contribution, shown in Section III, is qualitatively similar.
We find that SHC is reasonably high at some range of frequencies and momenta. 
Particulary interesting case is for static but spatially varying electric field:
The SHC resonates when the wavelength of the spatial variation of the electric field
matches with the spin precession length. A simple mechanism for this ``anomalous" spin Hall
current in 2DEG is described in Section IV. 
Section V is devoted for an experimental proposal to test this 
 novel mechanism, discussion and summary.

\section{Spin Hall Conductivity}

The spin Hall current for an electric field $\bm{E}(\q, \omega)$
at the wave vector $\q$ transverse to the direction of $\bm{E}$ and at the frequency $\omega$,
$\j_{y}^s(q, \omega) = \sigma_{yx}^s(q, \omega) E_x(q, \omega)$.
The spin Hall conductivity is nonlocal; spin current at a position $\r$ depends
on the electric field surrounding it: $\j_{y}^s(\r,\omega) =\int d\r' \sigma_{yx}^s(\vert
\r -\r'\vert, \omega) E_x(\r', \omega)$.
Using Kubo formula \cite{Mahan}, we find the transverse spin Hall conductivity 
\bea
\sigma_{yx}^s (\q , \omega) &=& \frac{1}{2\pi} \Re \,
      Tr\left[ \int \frac{d\k}{(2\pi)^2}
        \hat{j}^3_y(\k +\frac{\q}{2})\hat{\G}_\k^A(0) \right. \nonumber \\
    &\times  & \left.     \left\{ \hat{j}^0_x
        (\k +\frac{\q}{2}) + \hat{J}^0_x(\q, \omega)
       \right\} \hat{\G}_{\k +\q}^R(\omega )\right]
\label{Pi}
\eea
with
\be
\hat{J}^0_x = \frac{1}{m\tau} \int \frac{d\k'}{(2\pi)^2} \hat{\G}_{\k'}^A(0)
       \left\{ \hat{j}^0_x (\k' +\frac{\q}{2}) + \hat{J}^0_x
         \right\} \hat{\G}_{\k' +\q}^R(\omega )
\label{J0}
\ee
Here retarded (advanced) Greens function for an energy $\epsilon$
can be written as
\be
\hat{\G}_\k^{R,A}(\epsilon) = \frac{1}{2}
 \sum_{s=\pm} \frac{\sigma_0 + s (k_y \sigma_1
     -k_x\sigma_2)/\vert \k \vert}{\epsilon - \xi_\k^s \pm \frac{i}{2\tau}} \, . 
\ee
Equations (\ref{Pi}) and (\ref{J0}) together describe sum over infinite series
of ladder diagrams. 
We solve the matrix equation (\ref{J0}) numerically
and then using Eq.~(\ref{Pi}) we calculate $\sigma_{yx}^s
(q, \omega)$ when $\bm{E}\parallel \bm{e}_x$ and $\q \parallel \bm{e}_y$, i.e.,
transverse $\sigma_{yx}^s (q)$. 
For a system with Fermi energy 
$\epsilon_F$ and spin-splitting energy $2\lambda\k_F$ with $k_F$ being the
Fermi momentum, we choose two parameters
$\Delta = \epsilon_F\tau$ and $\delta = 2\lambda k_F\tau$ comparing with
disorder broadening $1/\tau$.

\begin{figure}
\vspace{-0.5cm}
\centerline{\hspace{-4.5cm}{\epsfysize=8cm\epsfbox{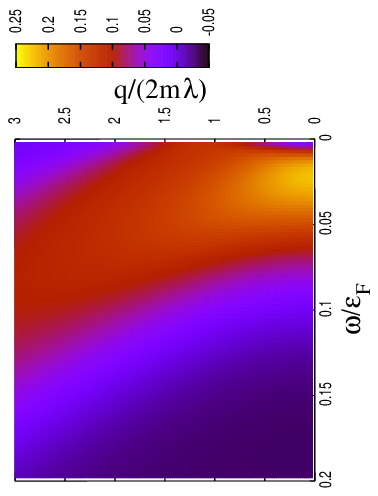}}}
\vspace{-8cm}
\centerline{\hspace{4.5cm}{\epsfysize=8cm\epsfbox{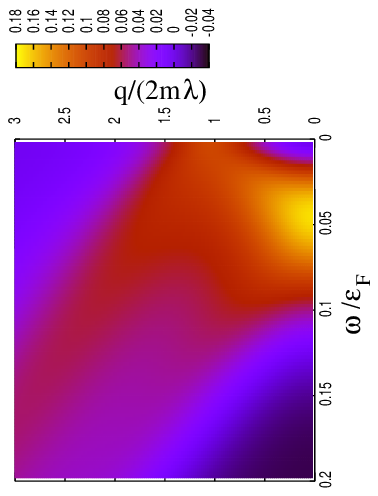}}}
\vspace{-1.5cm}
\centerline{\hspace{-4.5cm}{\epsfysize=8cm\epsfbox{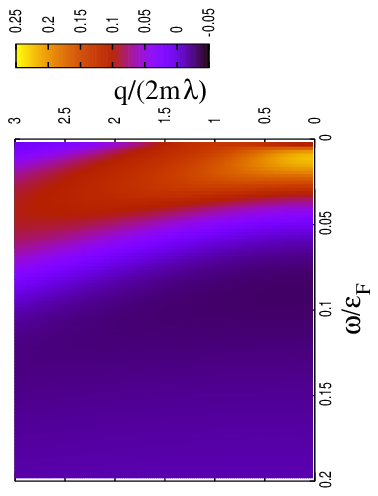}}}
\vspace{-8cm}
\centerline{\hspace{4.5cm}{\epsfysize=8cm\epsfbox{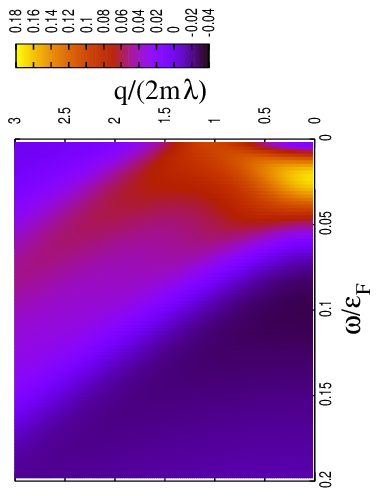}}}
\vspace{-1.5cm}
\caption{(Color online) Transverse spin Hall conductivity $\sigma_{yx}^s$ in the unit of $\delta^2(e/2\pi)$ 
as a function of $q/(2m\lambda)$ (horizontal axis) and $\omega/\epsilon_F$ (vertical axis).
The parameters $\Delta = 10,\, \delta = 0.4$ (upper-left panel), $\Delta = 20,\, \delta = 0.4$ (upper-right panel), 
$\Delta = 10,\, \delta = 0.8$ (lower-left panel), and $\Delta = 20,\, \delta = 0.8$ (lower-right panel) are considered.
}
\label{shall_omega}
\end{figure}

\begin{figure}
\vspace{-3cm}
\centerline{\hspace{0.5cm}{\epsfysize=13cm\epsfbox{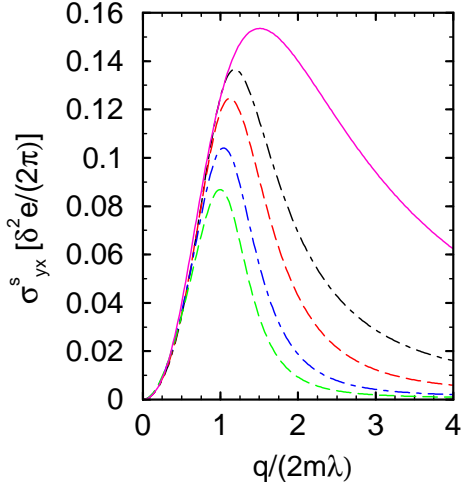}}}
\vspace{-4.0cm}
\caption{(Color online) Spin Hall conductivity $\sigma_{yx}^s$ (dot-dashed and dashed lines)
for the conventional definition
of the spin current {\it vs.} $q/(2m\lambda)$ for  
$\delta = 0.1, 0.4, 0.8 $ and $1.2$ from top to bottom
and for a fixed value of $\Delta =10$. 
$\sigma_{yx}^s$ is in the unit of $e\delta^2/2\pi$. $\sigma_{yx}^s$ is indeed
independent of $\Delta$ as we have checked for an wide range of $\Delta >1$. 
The maximum value of $\sigma_{yx}^s$ occurs at $q/(2m\lambda) \simeq 1$, 
The width of the peak in $\sigma_{yx}^s$ is larger
for smaller values of $\delta$ and the value of the peak
is larger for larger values of $\delta$. Solid line represents the analytical
expression (\ref{sigma}) of $\sigma_{yx}^s$ for small $\delta$.}
\label{shall}
\end{figure}

We show $\sigma_{yx}^s(q,\omega)$ for $\Delta = 10$ and $20$, and $\delta = 0.4$ and $0.8$ in figure \ref{shall_omega}.
The standard resonances occur at finite $\omega$ and at zero or very low value of $q$. The maximium value 
of $\sigma_{yx}^s(q,\omega)$ is almost proportional to $\delta^2$ and it decreases with the increase of $\Delta$. 
One common interesting feature for different combination of the parameters to notice is that the 
value of $\sigma_{yx}^s(q,\omega)$ is not small for $\omega \to 0$ and $q/(2m\lambda) \approx 1.0$.  
Figure \ref{shall} shows $\sigma_{yx}^s(q,0)$ for $\Delta =10$
and $\delta = 0.1,\, 0.4, \, 0.8$ and 1.2. 
These choices of $\delta$ correspond to $\ell < L_{so}$, where $L_{so} = \pi/(m\lambda)$ being the length 
traversed by an electron while its spin precesses by one cycle.
We have checked that $\sigma_{yx}^s(q,0)$ is
independent of $\Delta$ for an wide range of $\Delta >1$ while
$\delta$ is fixed and is almost proportional to $\delta^2$. 
 $\sigma_{yx}^s$ is zero at $q=0$ as we know from various calculations 
\cite{dis1,dis2,dis2a,dis3,dis4,dis5,num1,num2,sshall,Nagaosa}, 
and then it gradually increases with $q$ and 
form a peak around $q \simeq 2m\lambda$ before
it vanishes asymptotically. The position of the peak is almost
independent of $\delta$ but does depend on $\lambda$.
Since $q/(2m\lambda) = L_{so} /\Lambda$, the spin Hall current is maximum
when $\Lambda \simeq L_{so}$. In the limit of small disorder broadening,
i.e., for large $\delta$, $\sigma_{yx}^s(q)$ resonates exactly
at $q = 2m\lambda$ as in the case of $\delta=1.2$. 
Choosing different values of $\delta$ for
a fixed value of $\Delta$ implies different values of $L_{so}$. Larger
the value of $\delta$ means smaller $L_{so}$. As the value of
$L_{so}$ becomes smaller, the decrease in SHC will be faster from
its peak value for both increase and decrease of $\Lambda$.
This is the reason for narrower width of the SHC peak for 
larger values of $\delta$ as we see in Fig.~\ref{shall}.
The peak value of $\sigma_{yx}^s (q)$ is larger for larger $\delta$,
{\it i.e.} for larger $\lambda$ as well as $\tau$.

To demonstrate the resonance in $\sigma_{yx}^s(0,q)$ analytically, we calculate
$\hat{J}_x^0 $ in Eq.~(\ref{J0}) for $q_x=0$ and $q_y =q$. We sum over infinite
series of ladder diagrams starting with the contribution from ladder with
just one bar,
\be 
\frac{1}{m\tau} \int \frac{d\k'}{(2\pi)^2} \hat{\G}_{\k'}^A(0)
       \hat{j}^0_x (\k' +\frac{\q}{2}) \hat{\G}_{\k' +\q}^R(0 )
\equiv \sum_{\alpha =0}^3 J^0_\alpha \sigma_\alpha
\ee
Expressing $\hat{J}_x^0 = \sum_{\alpha =0}^3 {\cal J}_\alpha \sigma_\alpha$
and summing over geometrical series obtained from ladder diagrams we find
that only ${\cal J}_2$ and ${\cal J}_3$ survive at $q_x =0$ and they are
\bea
{\cal J}_2 &\approx & J^0_2\lb \frac{1}{1-I_{22}}+ \frac{I_{23}I_{32}}{
(1-I_{22})^2(1-I_{33})}\rb  \nonumber \\
 & &  +J_3^0\frac{I_{23}}{(1-I_{22})(1-I_{33})} \\
{\cal J}_3 &\approx& J^0_3 \frac{1}{1-I_{33}}
 +J_2^0\frac{I_{32}}{(1-I_{22})(1-I_{33})}
\label{J1}
\eea
where 
\be
I_{\alpha\beta} = \frac{1}{2m\tau}Tr\, 
\lb \int \frac{d\k'}{(2\pi)^2} \sigma_\alpha\hat{\G}_{\k'}^A(0)
       \sigma_\beta \hat{\G}_{\k' +\q}^R(0 ) \rb
\ee
Since the resonance occurs in $\sigma_{yx}^s(0,q)$ at $q\sim
2m\lambda \ll k_F$, we may wish to evaluate $I_{\alpha\beta}$ up to
quadratic in $q$ and hence the relevant components are $I_{22}
=1-(\Delta/m)\tau q^2-\delta^2/2$, $I_{33} = 1-(\Delta/m)\tau q^2 - \delta^2$,
$I_{32} = -I_{23} = 2i\delta\Delta q/k_F$ for $\delta \ll 1$. 
In this approximation, $J_2^0 = e\lambda\delta^2/2$ and $J_3^0 = -iqe\delta^2/4m$.
Using Eq.~(\ref{Pi}), we thus find
\be
\sigma_{yx}^s(0,q) \approx \frac{(e/2\pi)\delta^2 \tilde{q}^2}{(\tilde{q}^2+1)(
  \tilde{q}^2+2)}\lb \frac{5}{4} - \frac{1}{(\tilde{q}^2+1)}\rb
\label{sigma}
\ee
where $\tilde{q} = q/(2m\lambda)$. As we have seen in our numerical evaluation,
$\sigma_{yx}^s$ is independent of $\Delta$ and proportional to $\delta^2$ for
small $\delta$. The expression of $\sigma_{yx}^s$ (\ref{sigma}) is graphically
shown in Fig.~\ref{shall}. It agrees well with the numerical evaluation at low
$\tilde{q}$. The discrepancy at higher $\tilde{q}$ is expected as we have 
evaluated $I_{\alpha\beta}$, $J_2^0$ and $J_3^0$
analytically up to quadratic in $q$ only. 
Nevertheless the analytical expression (\ref{sigma}) explicitly shows the
resonance in $\sigma_{yx}^s (0,q)$.

\section{Spin Hall Current for Spin Torque}

\begin{figure}
\vspace{-0.5cm}
\centerline{\hspace{-4.5cm}{\epsfysize=8cm\epsfbox{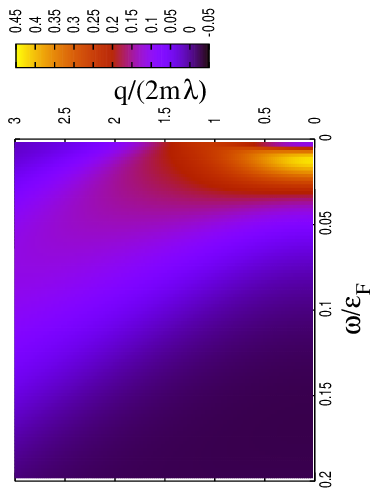}}}
\vspace{-8cm}
\centerline{\hspace{4.5cm}{\epsfysize=8cm\epsfbox{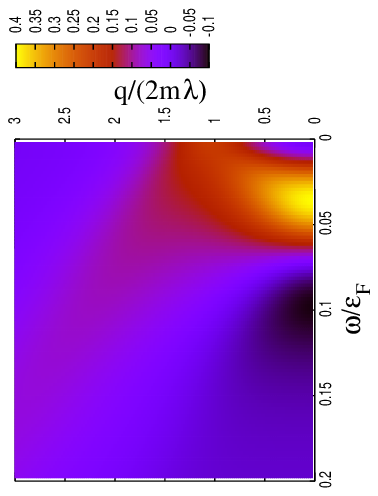}}}
\vspace{-1.5cm}
\centerline{\hspace{-4.5cm}{\epsfysize=8cm\epsfbox{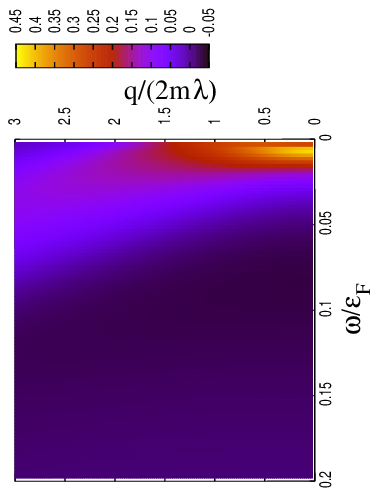}}}
\vspace{-8cm}
\centerline{\hspace{4.5cm}{\epsfysize=8cm\epsfbox{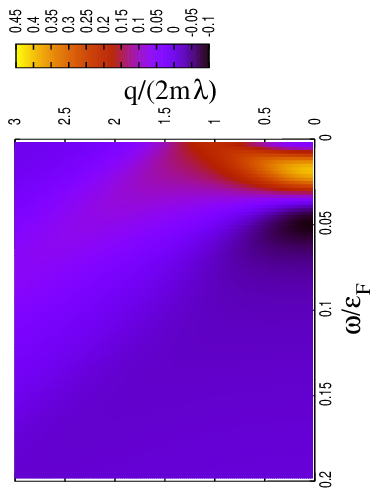}}}
\vspace{-1.5cm}
\caption{(Color online) Total transverse spin Hall conductivity $\sigma_{yx}^{cs}$ in the unit of $\delta^2(e/2\pi)$ 
as a function of $q/(2m\lambda )$ (horizontal axis) and $\omega/\epsilon_F$ (vertical axis).
The parameters $\Delta = 10,\, \delta = 0.4$ (upper-left panel), $\Delta = 20,\, \delta = 0.4$ (upper-right panel), 
$\Delta = 10,\, \delta = 0.8$ (lower-left panel), and $\Delta = 20,\, \delta = 0.8$ (lower-right panel) are considered.
}
\label{shall_total_omega}
\end{figure}

\begin{figure}
\vspace{-2.8cm}
\centerline{\hspace{0.5cm}{\epsfysize=13cm\epsfbox{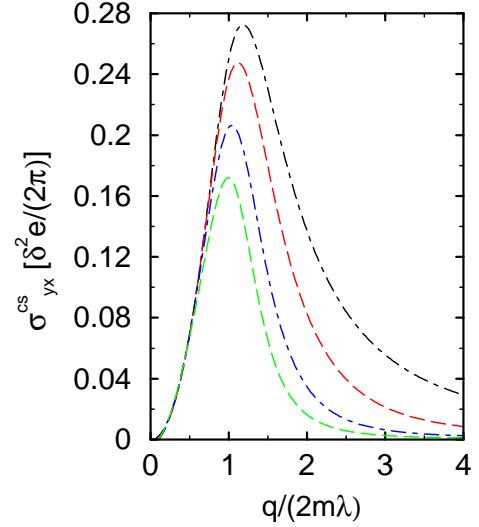}}}
\vspace{-3.6cm}
\caption{(Color online) Total spin Hall conductivity $\sigma_{sh}^{cs}(q)$ as a function
of $q/(2m\lambda )$. Other quantities and descriptions are 
same as in Fig.~\ref{shall}.}
\label{tshall}
\end{figure}

The above calculation of $\sigma_{yx}^s$ 
has been performed using the conventional definition of the 
spin current. We now consider the new definition of the spin 
current $\bm{{\cal J}}^3$
which is defined to satisfy continuity equation $\partial_t S_3
+ \bm{\nabla} \cdot \bm{{\cal J}}^3 = 0$ and can be expressed as the
sum of $\bm{J}^3$ and spin torque dipole density $\bm{P}^\tau$, {\it i.e.},
$\bm{{\cal J}}^3=\bm{J}^3+\bm{P}^\tau$ as proposed by 
Shi {\it et al} \cite{cspin}.  
Here $S_3$ and $\bm{J}^3$ are the spin density and conventional spin current
density operators respectively.
Further the spin torque density operator is expressed
as  $\tau_3(\r) = -\bm{\nabla} \cdot \bm{P}^\tau (\r)$ 
since the  average torque density vanishes in the bulk of the system. 
The second quantized form of the spin torque is
\be
\tau_3 (\q , t) =  \sum_{\k ,\alpha ,\beta} 
C_{\k ,\alpha}^\dagger(t) \hat{\tau}_{\alpha\beta} (\k +\frac{\q}{2})
  C_{\k +\q , \beta}(t)
\ee
where $C_{\k ,\alpha}^\dagger (t)$ is the electronic creation operator
of momentum $\k$ and spin $\alpha$ (up or down) at time $t$, and the spin
torque operator
$ \hat{\tau} (\k +\frac{\q}{2}) = \lambda (\k + \frac{\q}{2})\cdot \bm{\sigma}$.
We define spin torque--charge current correlation function
\bea
Q^{\tau 0}_x (q_x,q_y,\omega) &=& \frac{1}{2\pi}
      Tr\left[ \int \frac{d\k}{(2\pi)^2}
        \hat{\tau}(\k +\frac{\q}{2})\hat{\G}_\k^A(0) \right. \nonumber \\
      &\times & \left.  \left\{ \hat{j}^0_x
        (\k +\frac{\q}{2}) + \hat{J}^0_x(\q)
       \right\} \hat{\G}_{\k +\q}^R(\omega )\right]
\label{Q}
\eea
such that $\tau_3(0,q,\omega) = Q^{\tau 0}_x (0,q,\omega)E_x$.
Therefore the extra part of transverse spin Hall 
conductivity for an application of electric field
along $x$-direction is 
\be
\sigma_{yx}^{(2)}(q,\omega) =  -\frac{\Im Q^{\tau 0}_x (q,\omega)}{q}
\label{sigma2r}
\ee
The total (conserved) transverse SHC is then $\sigma_{yx}^{cs}(q,\omega)
=\sigma_{yx}^s(q,\omega)+ \sigma_{yx}^{(2)}(q,\omega)$. 
The values of $\sigma_{yx}^{cs}(q,\omega)$ for $\Delta =10$ and $20$, and $\delta = 0.4$ and $0.8$
are shown in Fig.~\ref{shall_total_omega}. The total transverse SHC is qualitatively similar to 
the same calculated for conventional spin current shown in Fig.~\ref{shall_omega}. 
Figure \ref{tshall} shows $ \sigma_{yx}^{cs}(q)$ at zero frequency
for $\Delta=10$ and $\delta =0.1,\,0.4,\,0.8,$ and $1.2$, although the
value of $\Delta$ is immaterial. The only relevant parameter is $\delta$.
The peak in SHC occurs at $q/(2m\lambda) \simeq 1$ and the width of the peak
is larger for smaller value of $\delta$ as in the previous case shown in Fig.~\ref{shall}.
Also the maximum value of $\sigma^{cs}_{yx}(q)$ is larger for larger $\delta$
and it is almost two times that of $\sigma_{yx}^s(q)$.

\section{Mechanism for anomalous spin Hall current}

\begin{figure}
\centerline{\epsfysize=6cm\epsfbox{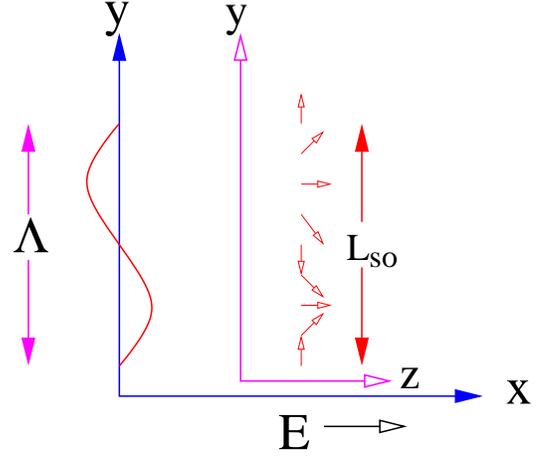}}
\caption{(Color online) Electric field along $x$-axis and its variation along $y$-axis
with wavelength $\Lambda$. Electrons moving along $y$-axis will have
spin precession in the $y-z$ plane. $L_{so}$ is the spin precession
length which is the length traversed by an electron while its spin precesses
by an angle $2\pi$. Lines with arrow indicate the direction of spin while it
precesses. The mode of rotation of spin changes when the sign of the electric
field changes. Therefore the spatial sign change of the electric field
induces a net out-of-plane spin current in the transverse direction. The spin
current will be maximum when $\Lambda \simeq L_{so}$. }
\label{cartoon}
\end{figure}

The time derivative of the charge current operator
is given by
\be
\partial_t  \hat{\j}_\k^0 = \frac{e}{m}\partial_t \k -4em\lambda^2 (\bm{e}_z \times \hat{\j}^3_\k) \, .
\ee
Therefore in the presence of external electric field $\bm{E}$,
the steady state equation becomes
\be
\frac{e^2}{m}\bm{E} 
= \frac{\langle \hat{\j}^0_\k\rangle}{\tau} + 4em\lambda^2 (\bm{e}_z \times \langle \hat{\j}^3_k\rangle )
\label{EOM}
\ee
where $\bm{e}_z$ is the unit vector along $z$-direction and angular brackets represent average
value of the quantity inside angular brackets. Therefore the electric
 field creates not only the charge current but also the spin current for
an electron with momentum $\k$. Although some electrons move perpendicular
to the electric field, their spin precess about $\Omega_\k = (k_y \bm{e}_x
-k_x\bm{e}_y)/\vert\k\vert$ while moving and hence no net spin
current flows in the bulk of the system. This is an alternative description
for vanishing $\sigma_{yx}^s$ in the system for static and uniform electric field. 
However the situation alters when
the electric field is static but nonuniform along its transverse direction. 
The basic physics behind the anomalous behavior for spatial
variation of the electric field is described in Fig.~\ref{cartoon}.
If the electric field is applied along $x$-direction, due to the wave
propagation along $y$-direction, sign of the electric field changes alternately
along $y$-direction with the wavelength $\Lambda = 2\pi/q$. 
The spin of an electron moving along $y$-direction will precess in the $y$--$z$
plane with the precession length $L_{so}$. 
However at the position where the sign of the electric field changes, the 
clockwise (anticlockwise) precession will change into anticlockwise (clockwise)
precession as in Fig.~\ref{cartoon}. This is because the sign of the $z$-component
of the spin changes as described by Eq.~(\ref{EOM}). 
The mode of spin precession for the electrons moving along
negative $y$-direction will be exactly opposite and therefore there will 
be a net $z$-polarized spin-current along $y$-direction. 
The spin current will be maximum when $L_{so} \simeq \Lambda$ and
it will sharply fall for the change in $\Lambda$ either way. 
This argument for {\em anomalous} spin hall current is somewhat similar to
Pippard's \cite{askin} ineffectiveness concept for anomalous skin effect 
in metals but the fundamental difference is that the former occurs for
$\Lambda \simeq L_{so}$ while the latter is due to               
$\Lambda \ll \ell$ ($\ell$ is the mean free path of an electron).
The present picture is ofcourse valid for $\ell < L_{so}$.

We have determined the  transverse wave number $q$ dependent spin Hall conductivity by
using Kubo formula in the previous two sections and show here 
how the basic physics presented in Fig.~\ref{cartoon} describes the {\em anomalous} spin Hall current.
On the other hand, there will be no longitudinal spin Hall current
 (when $\q \parallel \bm{E}$) for nonuniform but static electric field
because the electrons moving transverse to the electric field will not 
feel any change in sign of the electric field.

\section{Discussion and Summary}

\begin{figure}
\centerline{\hspace{0.5cm}{\epsfysize=4cm\epsfbox{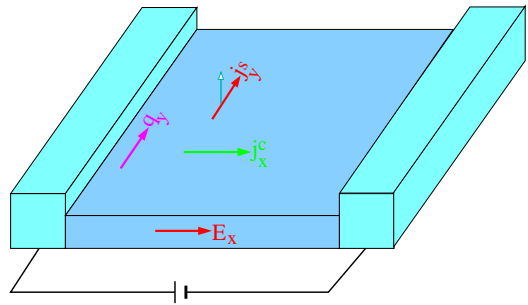}}}
\vspace{0.0cm}
\caption{(Color online) In a spin-Hall bar geometry, an electric current $j_x^c$ is driven
 through a 2DEG contacted with the metalic leads connected to a voltage source. The electric 
 field $E_x$ with transverse momentum $q_y$ induces a novel nonvanishing bulk spin-Hall current
 $j^s_y$ with spin pointing pependicular to the plane of 2DEG.}
\label{geometry}
\end{figure}

Spin accumulation observed by Sih {\it et al.} \cite{expt3}
in two dimensional electron
gas corresponds to the value of $\delta \sim 0.1$. 
This is in the large disorder
limit. Although the applied electric field is uniform, the electronic
inhomogeneity in the system due to disorder
may cause spatial variation of the electric field
in the system. The variation of the electric field describes the presence
of modes $\q$. If transverse $q$ is closer to $2m\lambda$, 
the contribution of these modes to the spin Hall conductivity is 
not negligible.
The presence of {\em anomalous} spin Hall current is then 
certain although the magnitude may be small since $\delta$ is small
in this experiment \cite{expt3}. 
This contribution is intrinsic because the spin orbit interaction is not 
disordered. Consideration of extrinsic mechanisms along with the intrinsic
spin orbit interaction provides finite \cite{intext1,intext2} 
spin Hall conductivity in presence of uniform electric field as well.    
However, any quantitative comparison of the {\em anomalous} spin Hall conductivity 
presented in this paper
with the experiment \cite{expt3}
or with the contribution arising from extrinsic effect \cite{intext1,intext2} 
or with the spin accumulation across the edges in a ballistic system \cite{Mishchenko}
is beyond the scope of the present study. 
Nevertheless our theory may be tested by applying a spatially varying
electric field with the variation along its transverse direction.
The geometry for an experimental proposal to test the novel mechanism for `anomalous'
spin Hall current is depicted in Fig.~\ref{geometry}.

The other studies at finite $q$ in this 
Rashba 2DEG are the induction of spin-density by electromagnetic wave 
\cite{Martin}, the response of the in-plane polarization 
\cite{Rashba3} to the transverse electric field in a pure system
and determination of density-density correlation function \cite{Mikhail} at all $q$.

In a cubic Rashba model which is relevant for two dimensional hole gas,
the intrinsic SHC is nonzero \cite{2dhg}, but the conserved spin Hall
conductivity vanishes \cite{Nagaosa} for short ranged impurity potential.
Therefore it is indeed interesting to look if these systems also have
anomalous spin Hall current \cite{future} like we have described here.

In summary, we have determined spin Hall conductivity at finite frequency
and finite {\em transverse} wavevector in 
 a disordered two dimensional electron gas with Rashba spin orbit interaction.
Interestingly at zero or small frequencies, we have found an {\em anomalous} spin Hall conductivity
which resonates when the wavelength of the spatial variation of the electric field matches
with the length of spin precession.
The mechanism responsible
for this is the change in the direction of spin precession for electrons
moving perpendicular to the electric field when the sign of the electric field
changes due to spatial variation. 
This is primarily due to the change in sign of the out-of-plane component of spin.

\acknowledgments{The authors thank J.K. Bhattacharjee, T.V. Ramakrishnan
 and K. Ray for stimulating discussions and M. Pletyukhov for sharing
his calculation before publication.}

\end{document}